\documentclass[a4paper,11pt]{article}
\usepackage{jinstpub} 

\usepackage{graphicx}
\usepackage{rotating}
\usepackage{supertabular}
\usepackage{longtable}
\usepackage{lscape}
\usepackage{multicol}
\usepackage{siunitx}
\usepackage{multirow}
\usepackage{bm}
\usepackage{adjustbox}
\usepackage{upgreek}
\usepackage{bm}
\usepackage{caption, subcaption}
\usepackage{wasysym}
\DeclareSIUnit\year{yr}

\title{Ultra-Low Background Germanium Assay at the Boulby Underground Laboratory}

\author{P. R.~Scovell$^{a,}\note{Corresponding author.}$, E.~Meehan$^a$, S. M.~Paling$^a$, M.~Thiesse$^b$, X.~Liu$^c$, C.~Ghag$^d$, M.~Ginsz$^e$, P.~Quirin$^e$, D.~Ralet$^e$}
 
 \affiliation[a]{STFC, Boulby Underground Laboratory, Boulby Mine, Redcar-and-Cleveland, TS13 4UZ, UK}
 \affiliation[b]{University of Sheffield, Department of Physics and Astronomy, Hounsfield Road, Sheffield, S3 7RH, UK}
 \affiliation[c]{SUPA, School of Physics and Astronomy, University of Edinburgh, Peter Guthrie Tait Road, Edinburgh, EH9 3FD, UK}
 \affiliation[d]{University College London (UCL), Department of Physics and Astronomy, Gower Street, London, WC1E 6BT, UK}
 \affiliation[e]{Mirion Technologies (Canberra) SAS ; 1 Chemin de la Roseraie, 67380 Lingolsheim, France}
 
\emailAdd{paul.scovell@stfc.ac.uk}

\abstract{
As we move to an era where next generation ultra-low background particle physics experiments begin to be designed and constructed, the ability to assay materials with high sensitivity and at speed with a variety of techniques will be key.

This paper describes the Mirion Technologies (Canberra) specialty ultra-low background detectors installed and commissioned at the Boulby Underground Laboratory between 2017 and 2021. The low background levels of the detectors combine with low background shielding and a radon-reduced dry nitrogen purge system to give sensitivity approaching the best in the world without the need for intricate shielding solutions. 

For an optimised sample geometry, run for \qty{100}{\day}, it would be possible to reach close to \qty{10}{\micro\becquerel\per\kg} ($10^{-12}$ g/g) for background radionuclides of interest in neutrinoless double-beta decay.
}

\keywords{Gamma detectors, Simulation methods and programs}

\arxivnumber{2308.03444} 

\begin{document}
\maketitle
\flushbottom

\section{Introduction}
Material radioassay for low-background particle physics experiments is of key importance. A detailed understanding of material radioactivity allows a precice calculation of the expected background signals over the duration of a run. Radioassay with germanium detectors has played a part in a number of low-background material characterisation campaigns in the areas of direct dark matter searches~\cite{LZ:2020fty,AKERIB2020102391, Aprile2017,Aprile2022,Aprile2015}, neutrinoless double-beta ($0\nu\beta\beta$) decay searches~\cite{ABGRALL201622,TSANG2023168477,10.1063/1.3579580}, neutrino experiments~\cite{Marti:2019dof,Abusleme2021,Fechner2011-ns} and others. 

Experiments looking for $0\nu\beta\beta$ decay are particularly concerned about radionuclides which have characteristic gamma-ray decays above the $Q_{\beta\beta}$ energy of the double-beta decay radionuclide of interest as the energy deposition associated with the Compton scattering of these gamma-rays may mask $0\nu\beta\beta$ events. This is also true for radionuclides with a $Q_{\beta}$ end-point energy greater than the $Q_{\beta\beta}$ energy of the double-beta decay radionuclide of interest. Of particular concern is $^{214}$Bi as it has a $Q_{\beta}$ end-point energy of \qty{3.27}{\mega\eV} which is above most $Q_{\beta\beta}$ energies.

The Boulby Underground Laboratory is located in the north-east of England at Boulby Mine. The laboratory is at a depth of \qty{1100}{\m} (\qty{2840}{\m} water equivalent) in a salt strata deposited some 250 million years ago when the Zechstein sea evaporated. The Boulby UnderGround Screening (BUGS) Facility has been operational since 2015 and has been important in the material characterisation efforts of several leading low-background particle physics experiments~\cite{LZ:2020fty,Marti:2019dof} and a number of environmental studies~\cite{Aguilar-Arevalo:2020alt,Aguilar-Arevalo:2020wfq,10.3389/fspas.2020.00050}. The original detectors which still operate in the facility have previously been discussed in depth~\cite{Scovell:2017srl}. In 2017, the facility began a process of upgrade to meet the material characterisation requirements of next generation dark matter and $0\nu\beta\beta$ decay experiments~\cite{Agostini:2022ido}. 

This process included the purchase of three specialty ultra-low background (S-ULB) detectors developed by Mirion Technologies (Canberra). Following the success of the original detectors, it was decided that the same philosophy should be employed to provide sensitivity from a few\,keV to \qty{3000}{\keV} to allowing characterisation of gamma-rays from $^{210}$Pb (\qty{46.5}{\keV}) through to $^{208}$Tl (\qty{2614.5}{\keV}). No single detector is suitable for maximising sensitivity across the whole energy range so a combination of Broad Energy Germanium (BEGe) and standard p-type coaxial germanium detectors were acquired. In addition, the Lumpsey SAGe-well detector was previously shown to have a background level which, while in specification, substantially reduced its potential sensitivity. This detector was returned to Mirion Technologies (Canberra) for refurbishment to the S-ULB standard. BUGS also operates two XIA UltraLo-1800 surface alpha counters, a dual detector low background radon emanation facility and an Agilent-8900 ICP mass spectrometer~\cite{DOBSON201825}. The addition of these assay techniques prompted the change in the acronym for BUGS (formerly the Boulby Underground Germanium Suite).

\section{Experimental Setup}
\subsection{The Detectors}

For the manufacture of S-ULB detectors, careful selection is made of all detector materials that will be located inside the lead shield. All materials close to the High Purity Germanium (HPGe) crystal are selected based on their activity to minimise the impact on the detector background. Screening of electronic components, and material such as aluminum and copper together with a complete control of all stages of the detector manufacturing process is key to the consistent and reproducible level of background. Furthermore, any material that is susceptible to cosmogenic activation is stored underground (800 meters water equivalent) to minimise the presence of potentially long half-life contaminants. This procedure was applied to \textit{Rosberry}, \textit{Belmont}, \textit{Merrybent} and \textit{Lumpsey}. There is no compromise between low radioactivity material and spectroscopic performance when compared to the standard version of each detector.

\begin{figure}[ht!]
\begin{subfigure}{.5\textwidth}
  \centering
  \includegraphics[width=1.\linewidth]{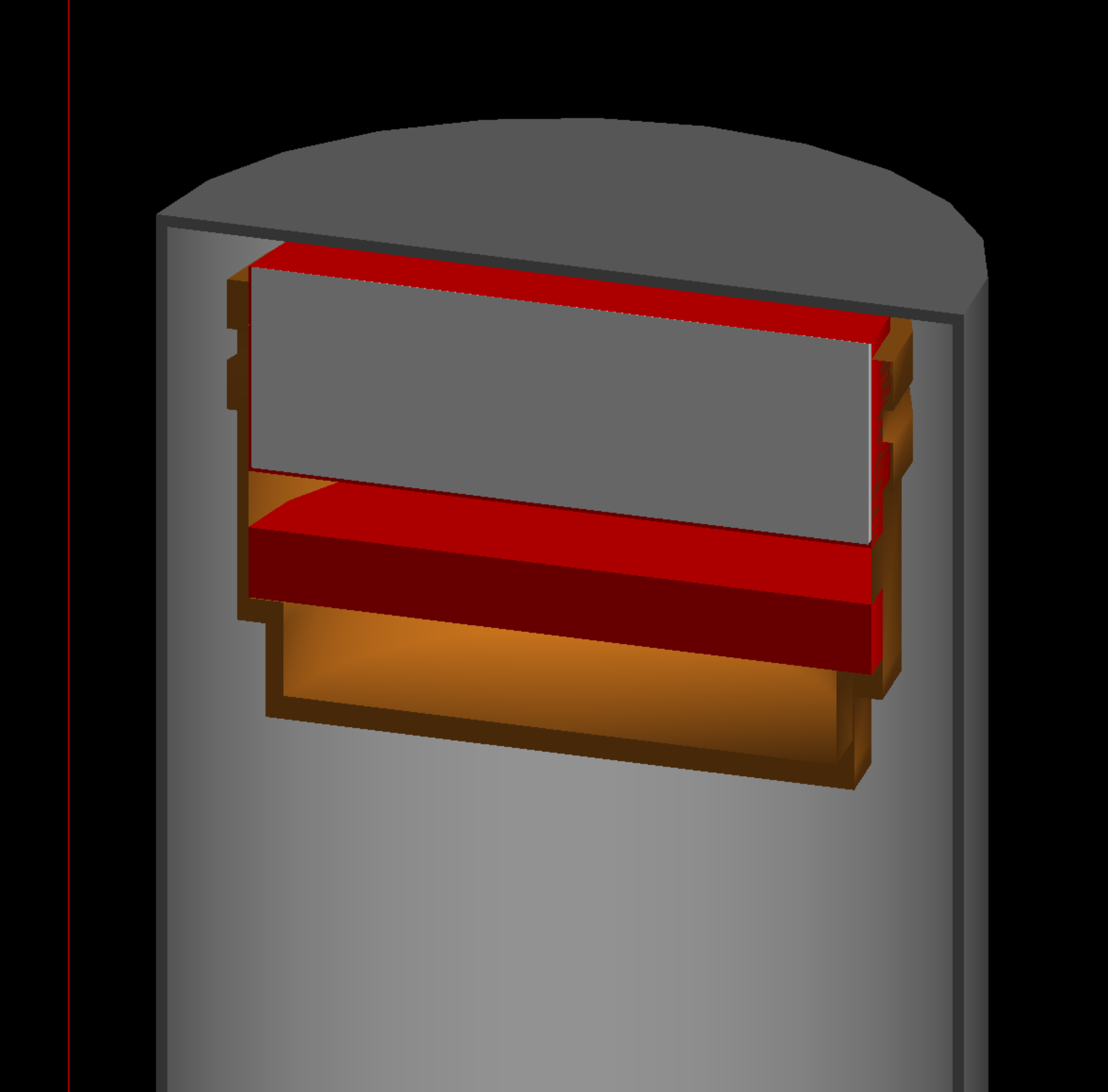}
  \label{fig:sfig1}
\end{subfigure}%
\begin{subfigure}{.5\textwidth}
  \centering
  \includegraphics[width=1.\linewidth]{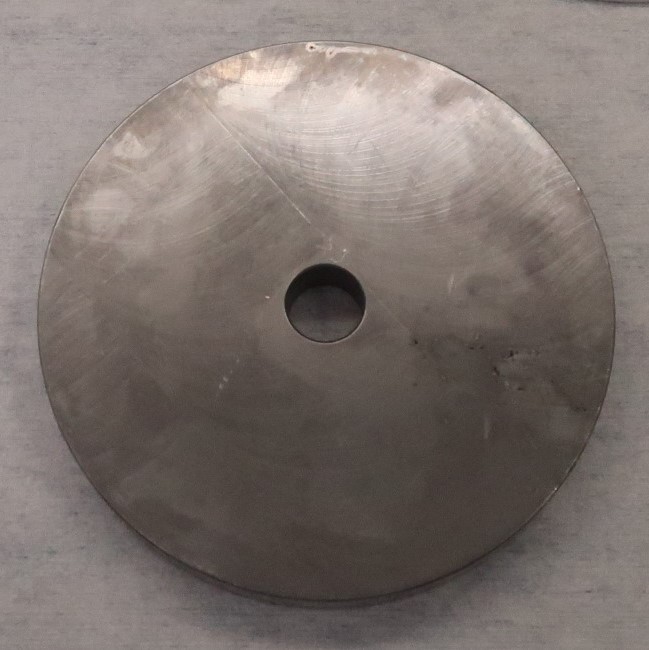}
  \label{fig:sfig2}
\end{subfigure}
\centering
\caption{\small{A GEANT4 rendering of the Roseberry BEGe detector (left). The HPGe crystal is shown in grey and the ultra-low background aluminium crystal holder is shown in orange. Behind the crystal (shown in red) is the ultra-low background lead disk which reduces any potential background from the back of the detector. The lead disk used is also shown in the picture on the right.}}
\label{fig:RoseberryLead}
\end{figure}

\noindent\textit{Roseberry} - Roseberry is a Mirion Technologies (Canberra) BEGe BE6530 detector. This planar detector has a \qty{65}{\cm\squared} area on its front face and a \qty{30}{\mm} thickness. The crystal is held in an ultra-low background cryostat and is shielded from any gamma-rays coming from the crystal holder and high voltage feedthrough by a thin disk of ultra-low background lead. 
In addition to selecting low-radioactivity material, the mass of material in the detection head is optimised such that it only contains essential components.
In addition, a thin outer contact is used on the HPGe to have a very thin ($\sim$\unit{\micro\meter}) residual dead-layer. This minimises low-energy gamma-ray absorption. Figure~\ref{fig:RoseberryLead} shows the internal configuration of the Roseberry detector with the ultra-low background lead disk visible.

\noindent\textit{Belmont} - Belmont is a \qty{600}{\cubic\cm} p-type coaxial high-purity germanium detector. A number of the materials used to construct the cryostat for this detector were assayed using the original BUGS germanium detectors to ensure as low a background as possible. With such a high relative efficiency, Belmont has the potential to assay samples with a precision and sensitivity far beyond that of any other detector in the BUGS germanium suite.

\noindent\textit{Merrybent} - Merrybent is a \qty{375}{\cubic\cm} p-type coaxial high-purity germanium detector. Although smaller than Belmont, this detector is still able to assay samples with high efficiency and has a background far lower and a performance far better than the only other equivalent detector, Lunehead~\cite{Scovell:2017srl}.

Both Belmont and Merrybent have large masses of HPGe (\qty{3.2}{\kg} and \qty{2.0}{kg}, respectively) but still provide excellent energy resolution meaning the size of the crystal does not compromise performance. In addition, freshly pulled high purity germanium crystals were used to limit the cosmogenic activation of the sensitive element of these two detectors. 

\begin{figure}[ht!]
\includegraphics[width=0.7\linewidth]{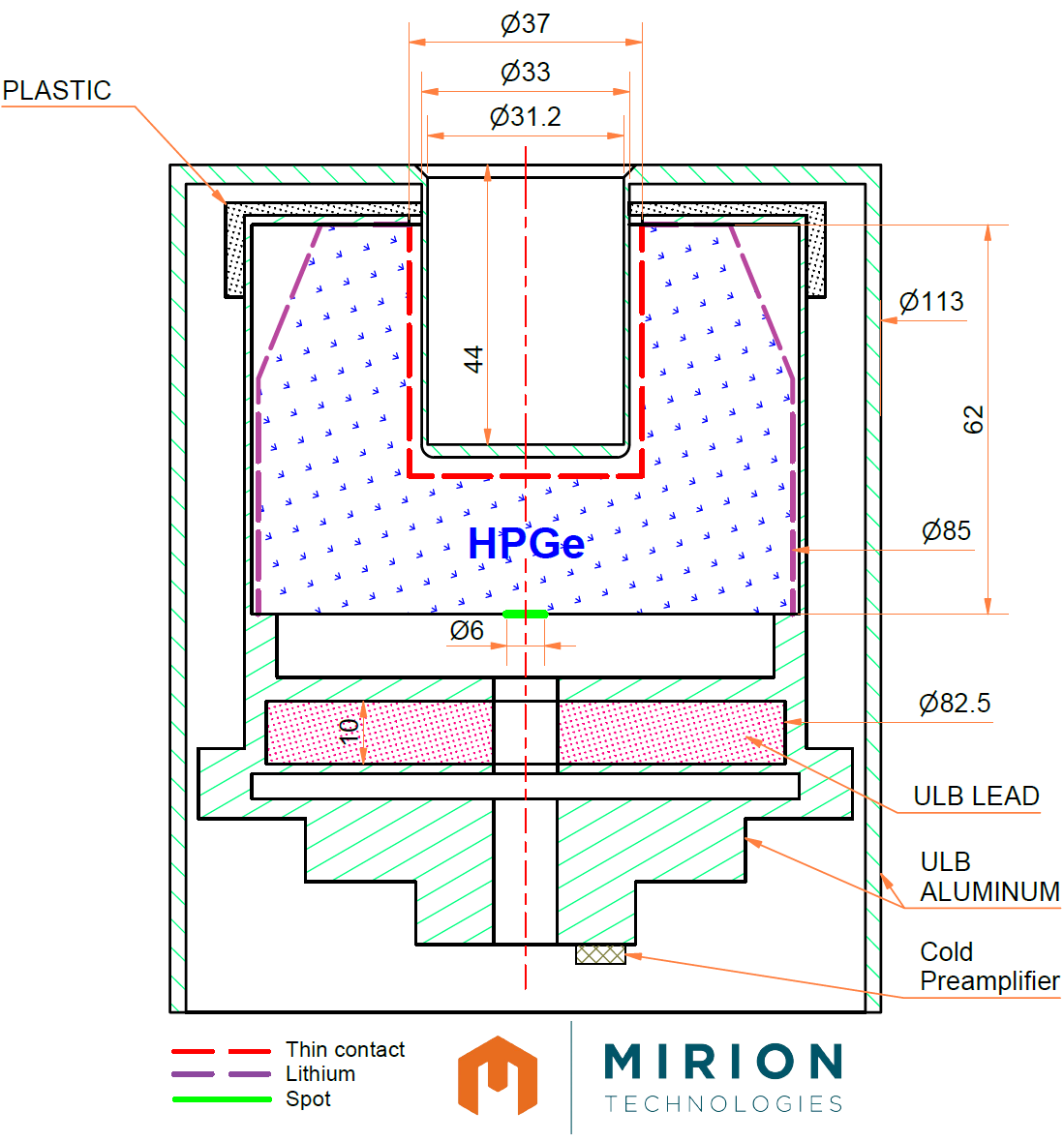}
\centering
\caption{\small{Internal drawing of the Lumpsey detector. Key dimensions are included (in mm) and major components are labelled. In this drawing, ULB means ultra-low background and the $\diameter$ symbol corresponds to the diameter.}}
\label{fig:Lumpsey_MCNP}
\end{figure}

\noindent\textit{Lumpsey} - Lumpsey is a SAGe well detector. Previously configured in a standard cryostat, it was not able to meet our sensitivity requirements for a number of materials. To maximise our sensitivity to low energy gamma-rays for small samples, this detector was refurbished to the S-ULB standard by Mirion Technologies (Canberra). The refurbishment of the detector followed the same standard as any S-ULB detector with a careful selection of material and electronics. As with any SAGe-Well detector, a particular focus was applied to select proper materials for the high voltage capacitor. In addition, the original configuration of this detector included a lithium drifted contact in the well. This left the detector susceptible to damage in the event of a warm up which could happen if access to the underground laboratory were to be restricted. The lithium contact was removed and replaced with a stable thin layer contact which minimises the absorption of low energy gamma-rays. This process involved increasing the diameter of the sample well in the germanium crystal from \qty{28}{\mm} to \qty{31.2}{\mm} meaning that samples \qty{\sim 20}{\percent} larger can be assayed when compared to the previous configuration. Finally, the refurbishment marks the first BUGS germanium detector to be electrically cooled. An internal schematic, modelled in MCNP, is shown Figure~\ref{fig:Lumpsey_MCNP} which shows the characteristic shape of the detector crystal.

\subsection{Shielding}

An important requirement of running the BUGS germanium detectors is to have a high level of confidence in the quality and stability of detector backgrounds. The first factor which determines this is the quality of the shielding material which surrounds the detectors. The shields for these detectors were produced by Lead Shield Engineering. All the detectors described above use multi-layer shielding with the inner layer comprising \qty{10}{\cm} of high-purity copper and the outer layer comprising \qty{10}{\cm} of lead. In addition, the shield surrounding the Belmont detector includes a \qty{10}{\mm} ultra-pure copper liner. As with the original BUGS germanium detectors, this lead and copper was sourced from stock which had previously been used to shield former low background experiments hosted at the Boulby Underground Laboratory.  

\subsection{Nitrogen Purge}

\begin{figure}[ht!]
\includegraphics[width=\linewidth]{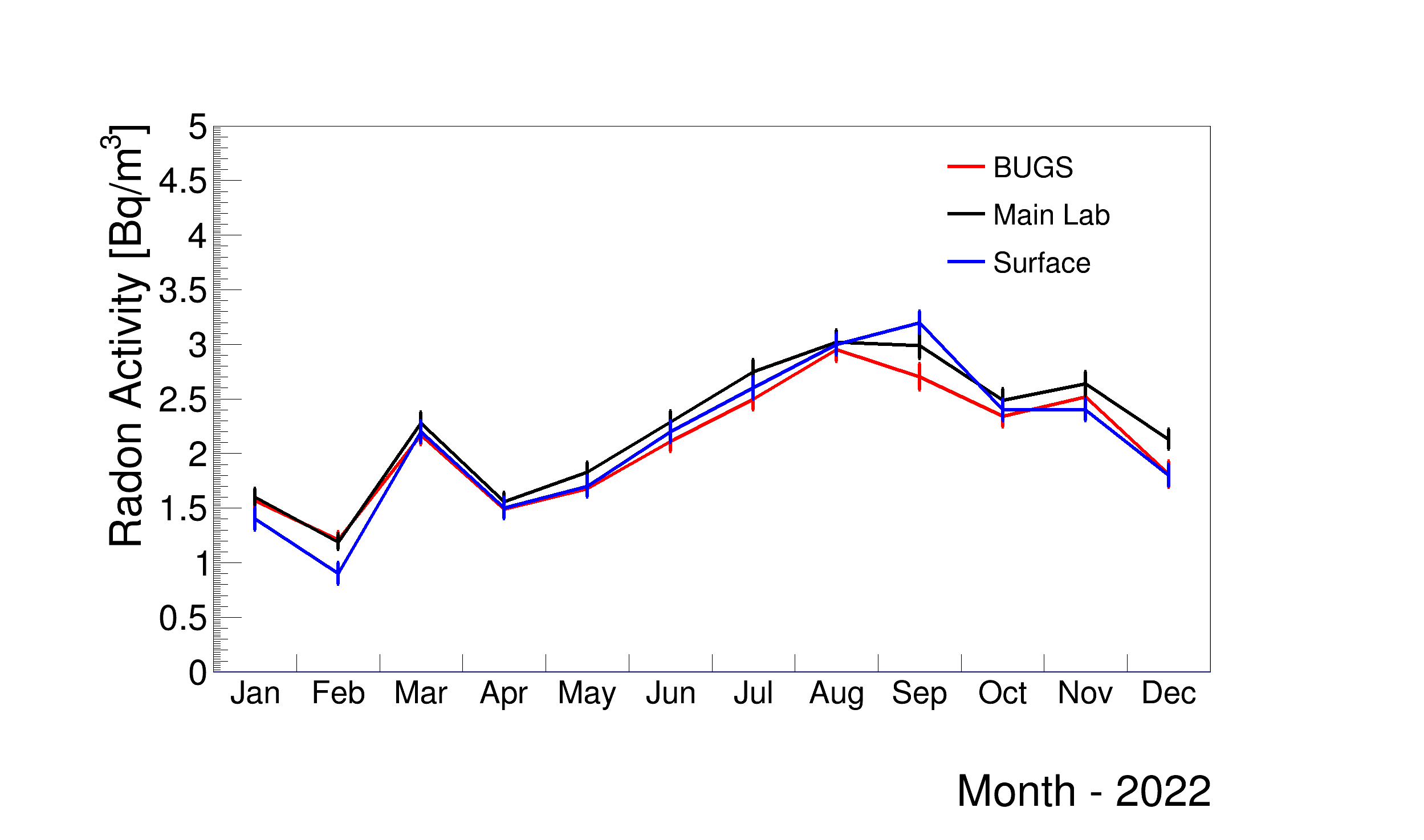}
\centering
\caption{\small{Monthly averages for radon activities in \unit{\becquerel\per\cubic\metre} for the surface, main Boulby Underground Laboratory and BUGS facility for 2022. It can be seen that the values are in reasonable agreement showing that there is no systematic increase in the radon levels underground as compared to the surface. The average values for the surface, main laboratory and BUGS are \qty{2.10+-0.03}{\becquerel\per\cubic\metre}, \qty{2.23+-0.03}{\becquerel\per\cubic\metre} and \qty{2.09+-0.03}{\becquerel\per\cubic\metre}, respectively.}}
\label{fig:radon}
\end{figure}

The second requirement for ensuring the quality and stability of the background in a germanium detector is to control the atmosphere which surrounds it.
The Boulby Mine ventilation system consists of a single intake down one of the access shafts. The air is circulated around the mine and then is exhausted up the second access shaft.
Figure~\ref{fig:radon} shows the monthly averages of radon detected both on the surface, in the main Boulby Underground Laboratory and in the BUGS facility during 2022. It can be seen that the underground radon levels mirror those seen on the surface. Over this measurement period, the average radon levels are \qty{2.10+-0.03}{\becquerel\per\cubic\metre}, \qty{2.23+-0.03}{\becquerel\per\cubic\metre} and \qty{2.09+-0.03}{\becquerel\per\cubic\metre} for the surface, main lab and BUGS, respectively. 

\begin{figure}[ht!]
\includegraphics[width=\linewidth]{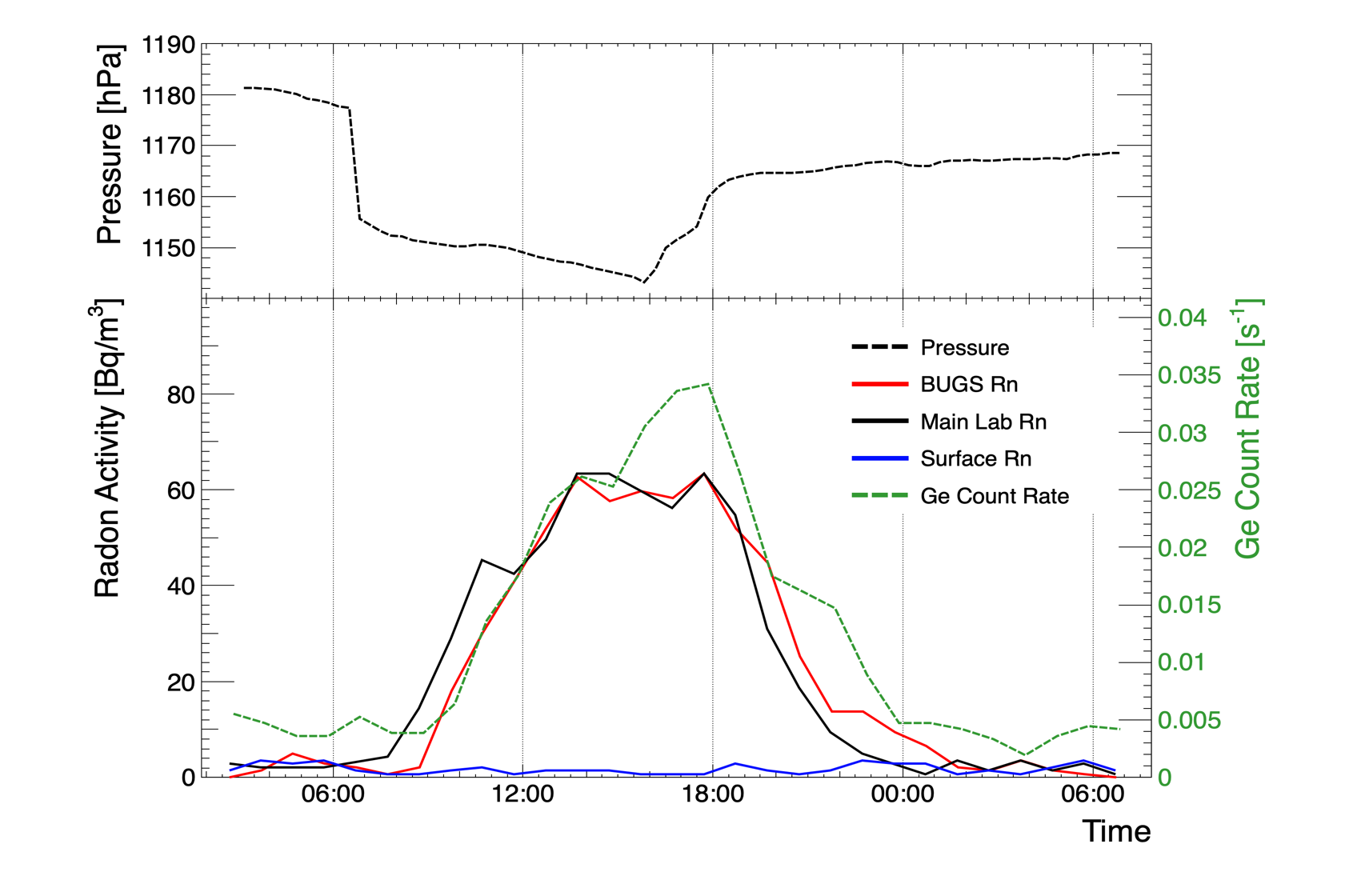}
\centering
\caption{\small{During routine mine ventilation fan maintenance, it can be seen that there is a substantial increase in radon levels in the underground laboratory (lower panel). The drop in pressure associated with the fans turning off (shown in the upper panel) allows air from deeper in the mine to travel back to the laboratory. This increase is mirrored in data from a sample running concurrently in the Roseberry detector. Acquisition is split into hourly slices so data with higher rates can be easily identified and removed.}}
\label{fig:radonSpike}
\end{figure}

This agreement shows that, with adequate ventilation, the radon levels measured underground are identical to those on the surface. This means that the radon emanation rate from material surrounding the laboratory is sufficiently low such that it is effectively mitigated by the intake ventilation air.
The rock surrounding the facility has relatively low levels of uranium when compared to similar underground facilities~\cite{Malczewski:2013lqy}. 
Daily averages show a variance of approximately \qty{1.1}{\becquerel\per\cubic\metre}. Figure~\ref{fig:radonSpike} shows the radon increase measured in the laboratory when the main ventilation fans for the mine are turned off for routine maintenance. It can be seen that the radon levels in the laboratory increase to approximately \qty{60}{\becquerel\per\cubic\metre}. This figure also shows the related increase in detector background for a sample running concurrently which peaks at approximately seven times higher than before and after the spike. Germanium data are taken in hour long snapshots so any periods with substantially higher radon levels can be identified and removed.

Any radon which enters the cavity of a germanium detector shield will cause unwanted background, particularly in spectral gamma-ray peaks associated with $^{214}$Pb and $^{214}$Bi which are products of the decay of $^{222}$Rn. To prevent the ingress of radon, a positive pressure of pure N$_2$ gas is used to purge the air from the cavity. This purge is often performed using boil-off from liquid nitrogen dewars. As the boiling point of radon (\qty{-61.7}{\degree C}) is substantially higher than that of liquid nitrogen (\qty{-195.8}{\degree C}), it stays liquid meaning that pure nitrogen is produced. In principle, this would be the ideal source of purge gas but in practice it can be difficult to maintain constant pressure and, therefore, flow rate.

Given the overall low levels of radon observed at the Boulby Underground Laboratory, we were confident that it would be possible to purge detectors with nitrogen gas generated on site. For the original BUGS germanium detectors, a simple system with a combination of Noblegen NG5 and NG6 nitrogen gas generators was used. This achieved acceptable results for the assay program for LZ~\cite{LZ:2020fty} but, as part of the upgrade, we wanted to ensure that the purest possible purge gas was available. Although the radon levels at Boulby are low, some impurities still remain in the initial nitrogen purge. To reduce this further, the gas is dried and passed through a charcoal trap held at \qty{-80}{\degree C}. To ensure that a constant flow of nitrogen is seen by each detector, the purge is controlled using Bronkhorst El-Flow Select mass-flow controllers which can flow at \qtyrange{0.2}{10.0}{\litre\per\minute} (\qty{+-0.6}{\percent} precision). This allows the identification of any periods with sub-optimal purge which, in turn, allows the removal of affected data.

\section{Characterising the Detectors}
\subsection{Backgrounds}
\begin{figure}[ht!]
\includegraphics[width=\linewidth]{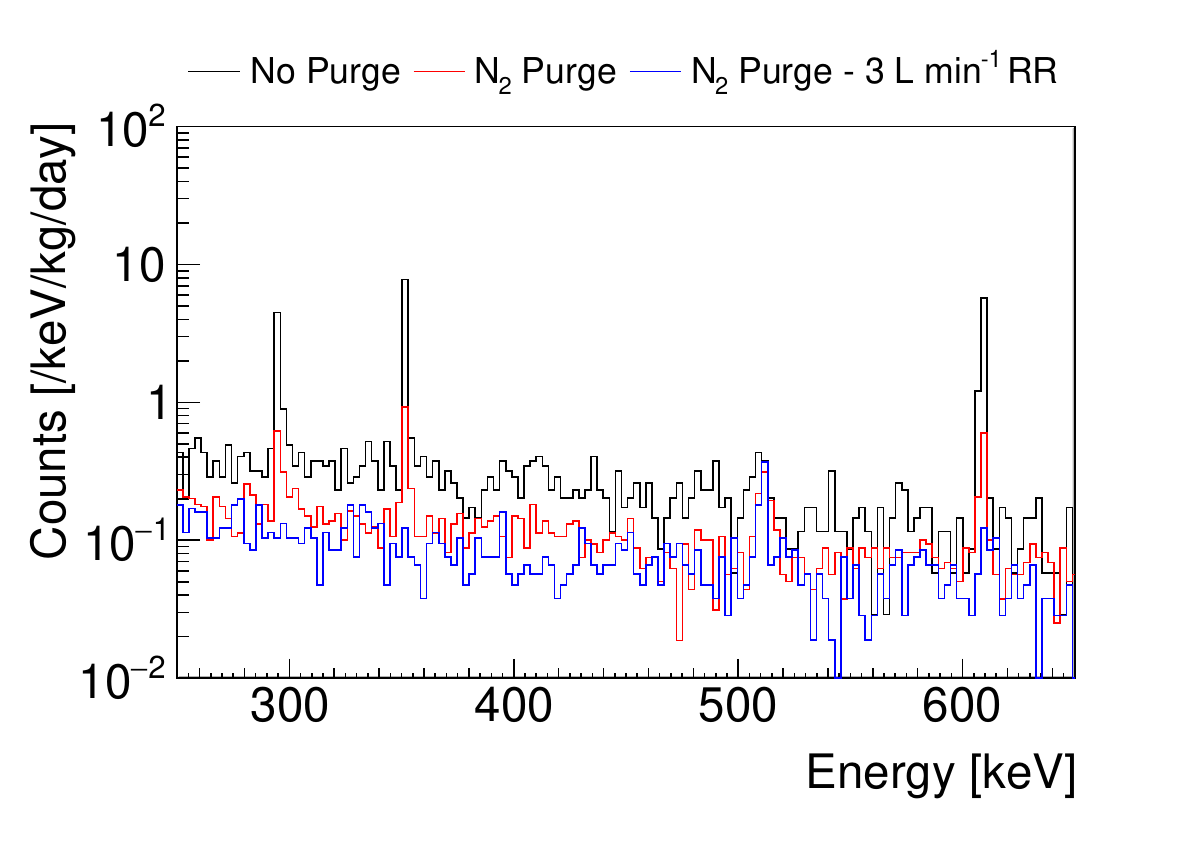}
\centering
\caption{\small{Comparison between the \qty{295}{\keV}, \qty{352}{\keV} and \qty{609}{\keV} sub $^{222}$Rn FEPs in the Belmont detector running with (black) no purge, (red) standard N$_{2}$ purge with no radon reduction and no MFC and (blue) the final purge setup with precisely \qty{3}{\litre\per\minute} of radon reduced N$_{2}$ delivered using an MFC.}}
\label{fig:Belmont_Bkg}
\end{figure}

\begin{figure}[ht!]
\includegraphics[width=\linewidth]{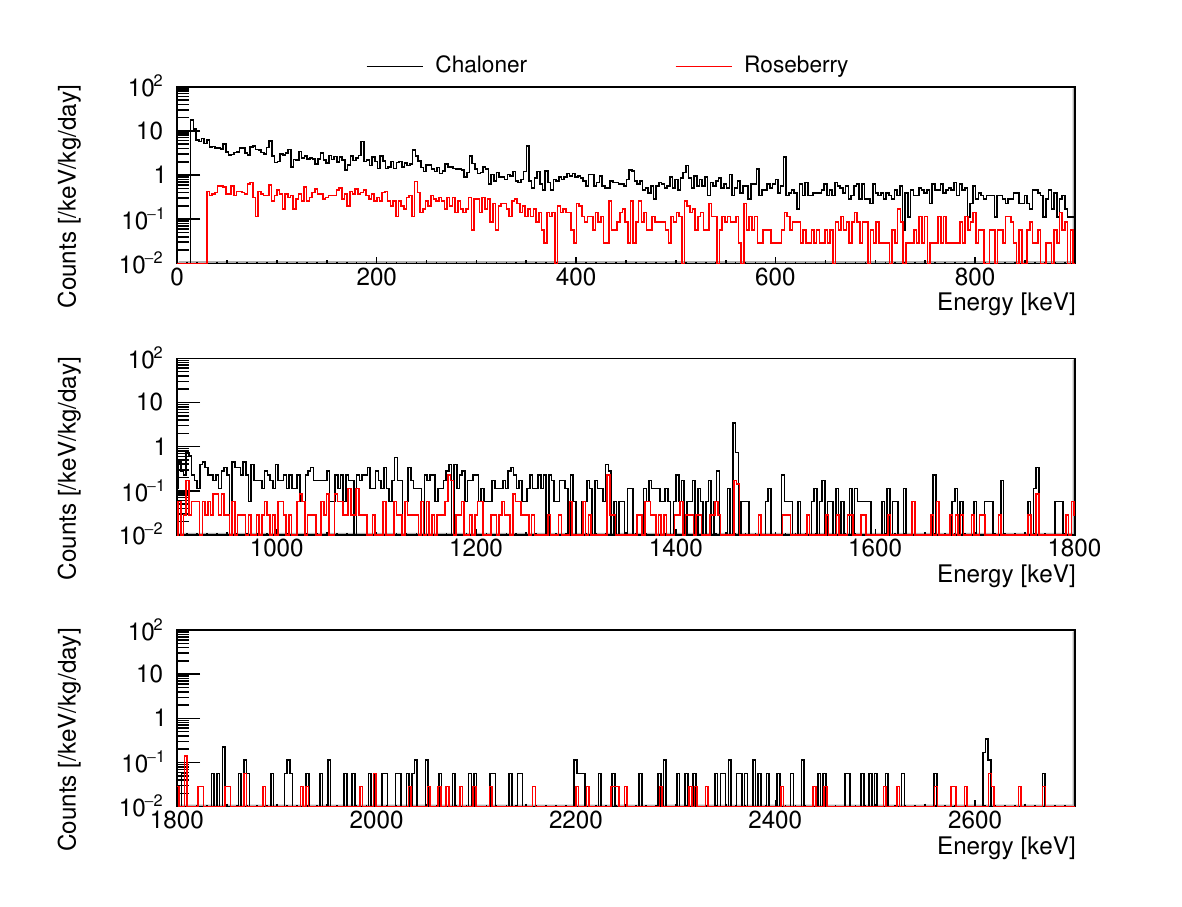}
\centering
\caption{\small{Comparison between the low background Chaloner (black) detector and the specialty ultra-low background detector Roseberry (red).}}
\label{fig:BEGeComparison}
\end{figure}

\begin{figure}[ht!]
\includegraphics[width=\linewidth]{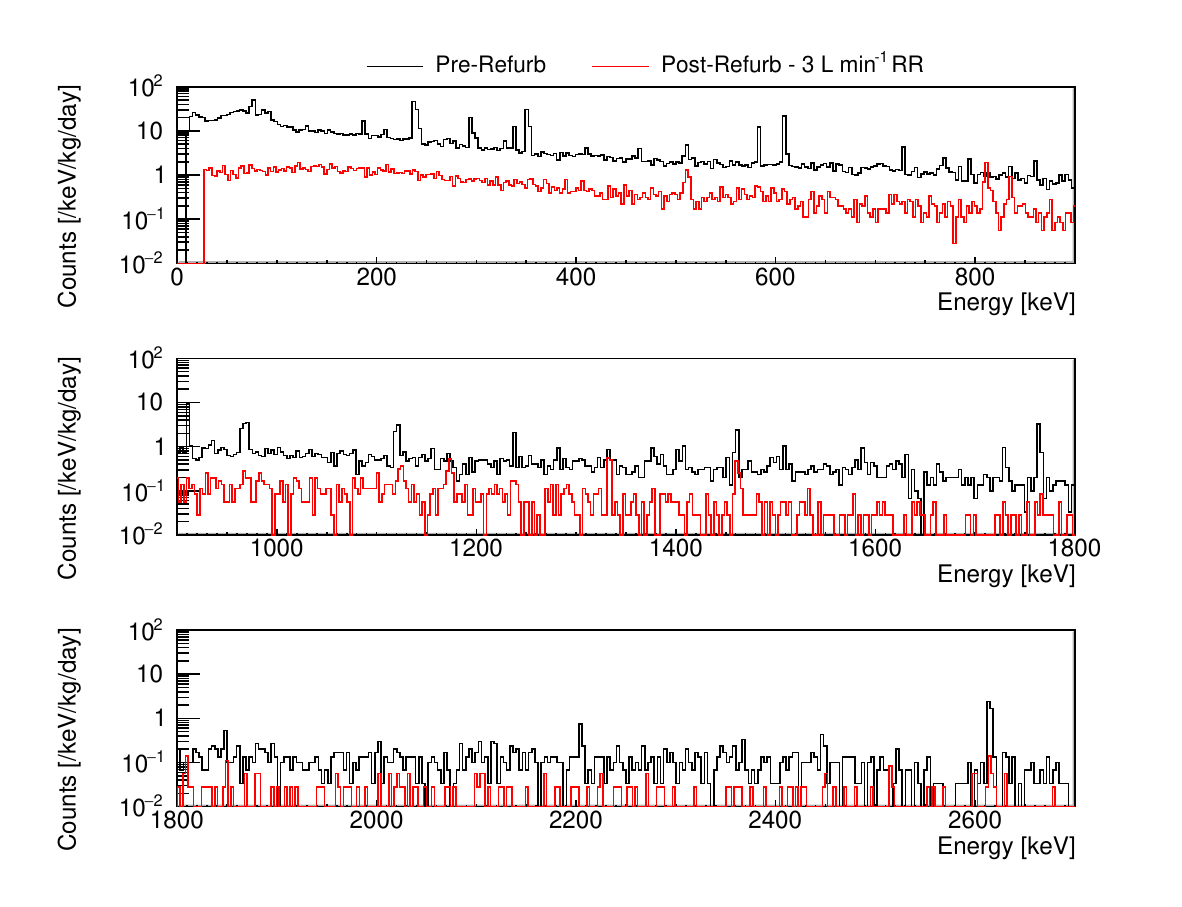}
\centering
\caption{\small{Comparison between Lumpsey running prior to refurbishment (black), and following refurbishment (red) with the final purge setup with precisely \qty{3}{\litre\per\minute} of N$_{2}$ delivered using an MFC. The overall background rate has reduced by a factor of 70 times but some individual peaks have reduced to far lower levels (see Table~\ref{tab:rates}). The FEPs visible at \qty{511}{\keV}, \qty{811}{\keV}, \qty{834}{\keV}, \qty{1173}{\keV} and \qty{1332}{\keV} are due to cosmogenic activation of the copper crystal holder as discussed in~\cite{Baudis:2015ptm}.}
\label{fig:Lumpsey_Bkg}}
\end{figure}

In early 2021, the detectors were running with the nitrogen purge operating in a stable and monitored condition with no additional improvements planned. By means of comparison, Figure~\ref{fig:Belmont_Bkg} shows the background levels for standard sub-$^{222}$Rn full energy peaks (FEPs) in the Belmont detector with three configurations. Firstly shielded but with no purge, secondly with the previous purge configuration (no radon reduction and minimal flow control) and finally using the radon reduced nitrogen flowing at \qty{3}{\litre\per\minute} through a mass flow controller (MFC). Figure~\ref{fig:BEGeComparison} shows a comparison between the backgrounds measured in Chaloner (a standard BE530 detector) and Roseberry. As expected, the background in Roseberry is substantially below that measured in Chaloner even with identical radon-reduced nitrogen purges. Finally, a comparison between Lumpsey pre- and post-refurbishment is presented in Figure~\ref{fig:Lumpsey_Bkg}. In its previous configuration, the background was high enough that the nitrogen purge made no appreciable difference to the measured background. We see that the Lumpsey background in the new configuration has a substantially lower background, with an overall 70 times reduction.
Table~\ref{tab:rates} summarises the count rates for several gamma-ray peaks in the detectors presented in this study and in the original BUGS germanium detectors~\cite{Scovell:2017srl} now running with the optimised radon-reduced nitrogen purge. The background counts in Chaloner and Lunehead are dominated by impurities in the detector and shield construction rather than from airborne radon, so are not affected by improvements in the N$_2$ purge. A measurement for Lumpsey in its previous configuration is also presented, showing the substantial reduction in background rate.

In its current configuration, Merrybent is not achieving the lowest background possible for late chain $^{238}$U due to what is believed to be contamination in the shield. Several assays with large, dense, and radiopure materials give counts in the $^{214}$Pb and $^{214}$Bi peaks at rates below the background rate~\cite{THIESSE2022110384}. This infers that the elevated background does not come from the detector and the performance of the N$_{2}$ purge in Roseberry and Belmont (noting again that all detectors share a common source of N$_{2}$ for the purge) infers that there should not be an issue with the purity of the gas. The lowest rates seen in this detector occurred during assay of ultra-pure gadolinium sulphate where count rates of \qty{1.1}{\per\kg\per\day} and \qty{0.6}{\per\kg\per\day} were observed for the \qty{351}{\keV} and \qty{609}{\keV} FEPs, respectively.

\begin{table}[ht!]
\centering
\caption{Count rates normalised to germanium crystal mass for the Boulby HPGe detectors in their current configuration. These values may not represent the ultimate backgrounds that could be achieved with these detectors but this is for the very simple shield and purge configuration described in this paper. With the exception of the pre-refurbishment Lumpsey background, these runs were all performed in early 2021. The values for Lunehead, Chaloner and Lumpsey (pre-refurbishment) are broadly in agreement with the measurements discussed in~\cite{Scovell:2017srl}.\label{tab:rates}}
\begin{adjustbox}{width=\textwidth,center}
\begin{tabular}{|l|c|c|c|c|c|c|c|}
\hline
\multirow{3}{*}{{Detector}} & \multicolumn{7}{c|}{{Count Rate ({\boldmath \unit{\per\kg\per\day}})}}\\
\cline{2-8}
&{Integral} & {351 keV} & {609 keV} & {238 keV} & {1461 keV} & {2615 keV} & {46.5 keV}\\
& {100--2700 keV}&  {$^{214}$Pb} & {$^{214}$Bi} & {$^{212}$Pb} & {$^{40}$K} & {$^{208}$Tl} &{$^{210}$Pb}\\
\hline \hline
Belmont & 90(9) & 0.2(1) & 0.4(2) & 0.13(8) & 1.0(2) & 0.3(1) & - \\
Merrybent & 145(12) & 2.5(3) & 1.8(3) & 0.3(1) & 1.9(3) & 0.8(2) & - \\
Lunehead & 540(25) & 5.6(5) & 4.7(4) & 8.3(5) & 9.1(6) & 2.0(3) & - \\
Roseberry & 130(11) & 0.15(7) & 0.15(7) & 0.8(3) & 0.8(2) & 0.2(1) & 0.4(6)\\
Chaloner & 1045(30) & 5(1) & 4(1) & 7(1) & 8.4(14) & 2.1(5) & 1.8(11)\\
Lumpsey - 2021 & 515(25) & 1.1(7) & 1.3(3) & 1.1(7) & 1.7(7) & 0.2(2) & 1.7(6) \\
Lumpsey - 2019 & 36880(6) & 114(4) & 68(3) & 172(5) & 8(1) & 11(1) & 14(2) \\
\hline
\end{tabular}
\end{adjustbox}
\end{table}

\subsection{Calibration}

The detectors are periodically calibrated using a check-source which includes \qty{37}{\kilo\becquerel} of $^{155}$Eu and \qty{37}{\kilo\becquerel} of $^{22}$Na. This serves to monitor the stability of peak resolution and can be used to monitor dead-layer thickness in the Roseberry BEGe detector. For the former, peak widths are extracted from gamma-ray peaks at \qty{86.5}{\keV} and \qty{105.3}{\keV} from $^{155}$Eu and at \qty{1274.5}{\keV} from $^{22}$Na. The \qty{511}{\keV} e$^{+}$e$^{-}$ annihilation peak from $^{22}$Na may also be used although Doppler broadening means that this is not useful to monitor absolute resolution. To monitor the stability of the dead layer, the ratio of the count rate in the \qty{86.5}{\keV} and \qty{105.3}{\keV} peak areas is used. If the dead layer begins to thicken, the area of the lower energy peak would reduce relative to the higher energy peak. A similar technique using a collimated $^{241}$Am source is described in~\cite{Zeng:2016mef} although we only monitor the overall dead-layer thickness, not the position dependent thickness. 

\begin{figure}[ht!]
\centering
\begin{subfigure}[t]{0.95\textwidth}
\centering
\includegraphics[width=0.65\textwidth]{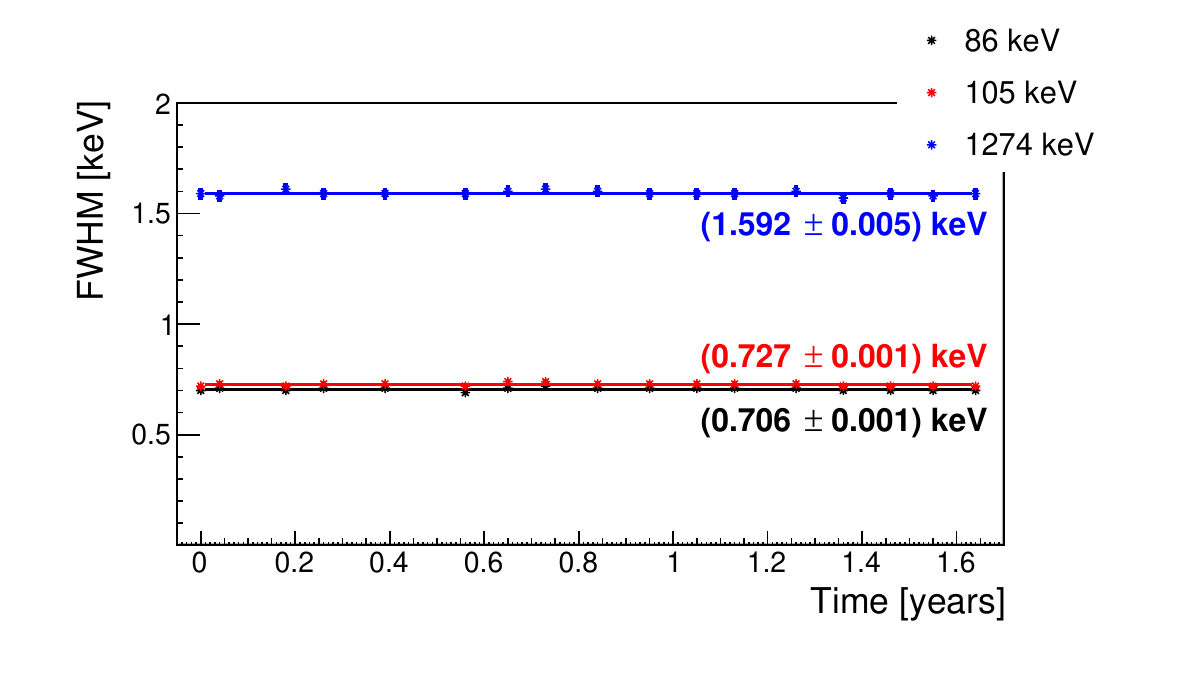} 
\caption{FWHM resolution measured for calibration FEPs on the Roseberry detector between March 2021 and November 2022. Statistical uncertainties are included but are not visible on this scale.} \label{fig:resn}
\end{subfigure}

\begin{subfigure}[t]{0.95\textwidth}
\centering
\includegraphics[width=0.65\textwidth]{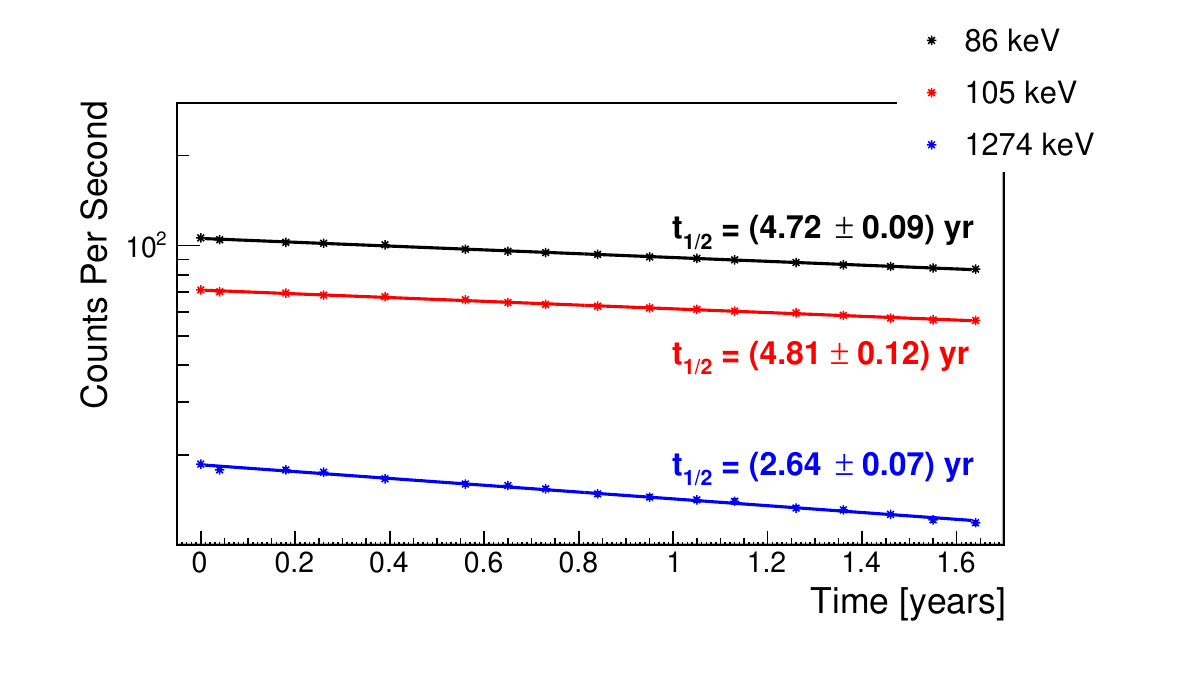} 
\caption{Counts per second for calibration FEPs on the Roseberry detector over the same period as seen in the figure above. The calculated half-lives are consistent with those for $^{155}$Eu (\qty{4.76}{\year}) and $^{22}$Na (\qty{2.60}{\year}). Statistical uncertainties are included but are not visible on this scale.} \label{fig:cps}
\end{subfigure}

\begin{subfigure}[t]{0.95\textwidth}
\centering
\includegraphics[width=0.65\textwidth]{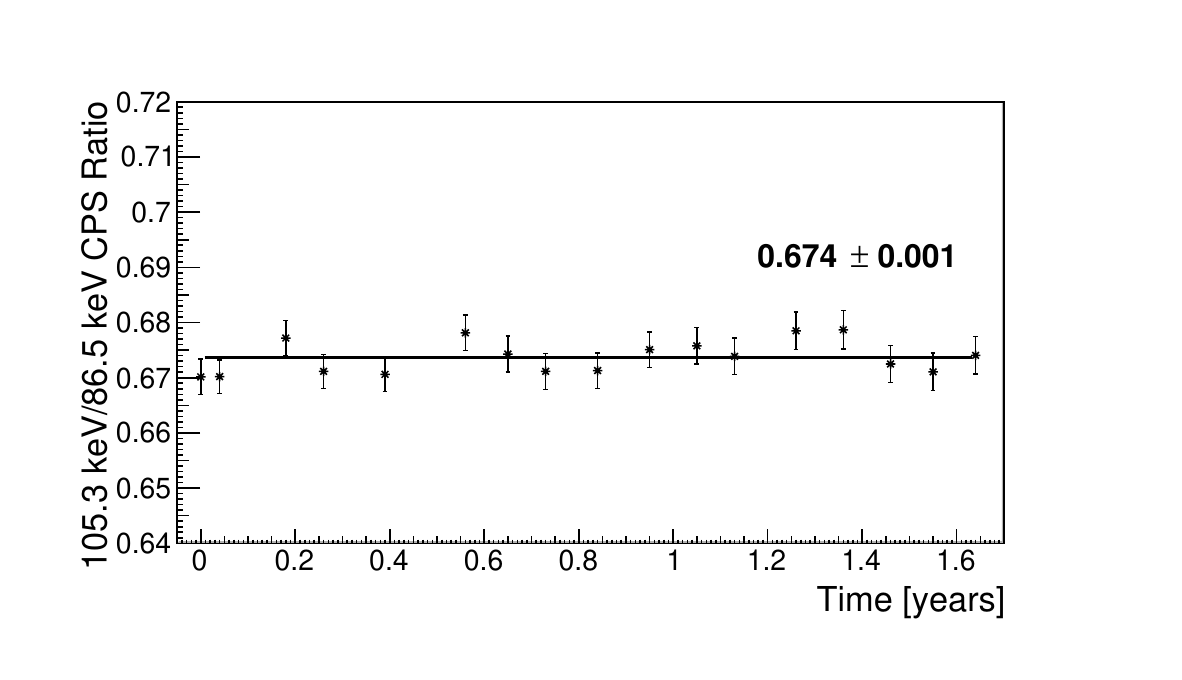} 
\caption{Ratio of the \qty{105.3}{\keV} and \qty{86.5}{\keV} FEPs from decay of the $^{155}$Eu radionuclide in the multi-gamma calibration source.} \label{fig:ratio}
 \end{subfigure}

 \caption{Calibration plots for a multi-radionuclide ($^{155}$Eu and $^{22}$Na) calibration source on the Roseberry detector.}

\end{figure}

Figure~\ref{fig:resn} shows the stability of FWHM resolution for the Roseberry detector for the calibration energies. Figure~\ref{fig:cps} shows the count rates over time, demonstrating the agreement between the decrease in counts per second and the half lives of $^{155}$Eu and $^{22}$Na which are \qty{4.76}{\year}~\cite{NICA20191} and \qty{2.60}{\year}~\cite{BASUNIA201569}, respectively. Figure~\ref{fig:ratio} shows the stability of the 105.3/86.5 peak area ratio over a period of \qty{1.6}{\year} from March 2021 to November 2022. This gives a constant ratio of $0.674\pm0.001$ with a reduced chi-squared of 0.87.

\subsection{Geometric Efficiency and Minimum Detectable Activity}
To provide a comparison between detectors, we employ a similar technique to determine geometric efficiency as described in~\cite{Scovell:2017srl}. For the larger p-type detectors, sensitivity is maximised when the combination of sample mass and efficiency is maximised. For Belmont and Merrybent, the assay of a powder filled \qty{4}{\litre} Marinelli beaker (\qty{4.5}{\kg}) is simulated. This is about the largest sample these detectors would encounter. The density of the powder varies depending on sample size and the elemental composition is shown in Table~\ref{tab:powder}. The geometric efficiency of HPGe detectors had been studied in depth using a number of techniques~\cite{Venkataraman,BRITTON201512,BARRIENTOS2011S228} but in this case we use a Monte-Carlo simulation constructed using GEANT4~\cite{AGOSTINELLI2003250}.

\begin{table}[ht!]
\centering
\caption{Elemental composition of the powder used in the MDA simulaton. Taken from the Mirion (Canberra) geometry composer software~\cite{ICRU:1994}.  \label{tab:powder}}
\begin{tabular}{|c|c|c|c|c|}
\hline
Element & Composition (\unit{\percent}) \\
\hline
Oxygen & 56 \\
Silicon & 32 \\
Aluminium & 7 \\
Iron & 3 \\
Carbon & 1\\
Hydrogen & 1\\
\hline
\end{tabular}
\end{table}

\begin{figure}[ht!]
\includegraphics[width=\linewidth]{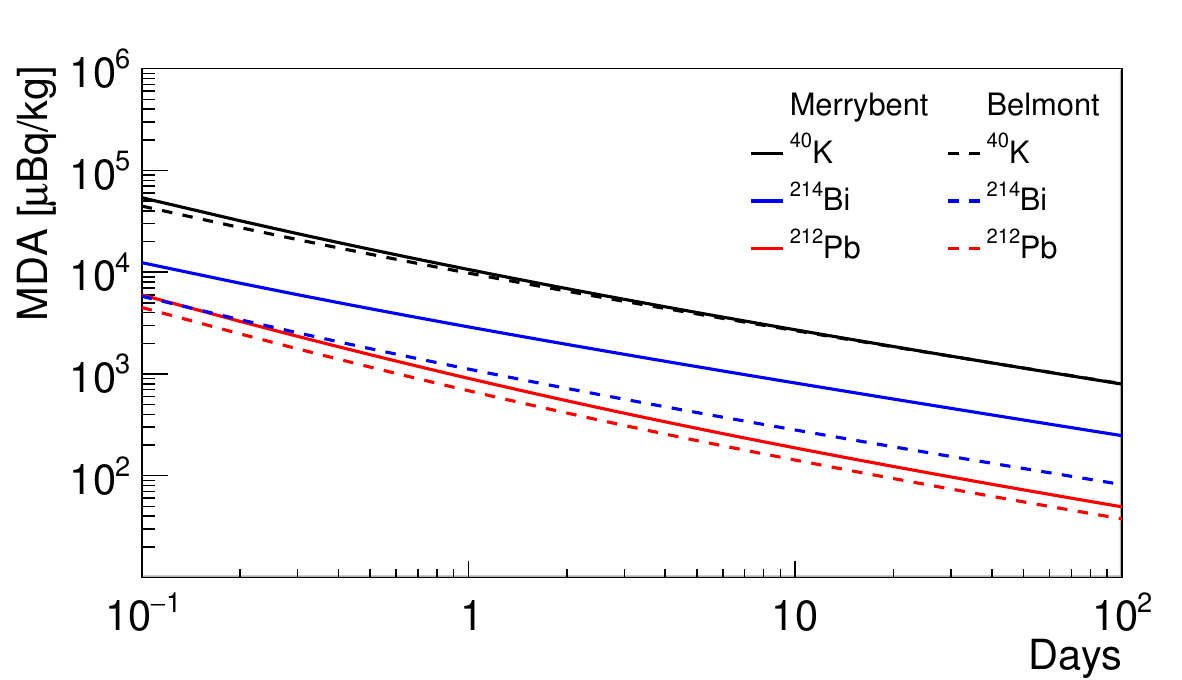}
\centering
\caption{\small{Comparison between the MDAs for the radionuclides listed in Table
~\ref{tab:mda_0nubb} on the Merrybent and Belmont detectors. It can be seen that the sensitivity is generally better in Belmont with the exception of $^{40}K$ where a lower count rate in Merrybent (despite a lower efficiency) for this FEP, leads to a better predicted result.}}
\label{fig:belmer_mda}
\end{figure}

In Figure~\ref{fig:belmer_mda}, it can be seen that the best results are for the $^{212}$Pb radionuclide which forms part of the $^{232}$Th chain. Due to the short half lives of $^{212}$Pb (10.6 hr), $^{212}$Bi (61 min) and $^{208}$Tl (3.1 min), we can assume that these radionuclides are in secular equilibrium and, as such, assume that a measure of the \qty{238}{\keV} FEP from $^{212}$Pb will give a good prediction for the \qty{2614.5}{\keV} FEP from $^{208}$Tl.

\begin{figure}[ht!]
    \centering
    \begin{minipage}{0.45\textwidth}
        \centering
        \includegraphics[width=0.9\textwidth]{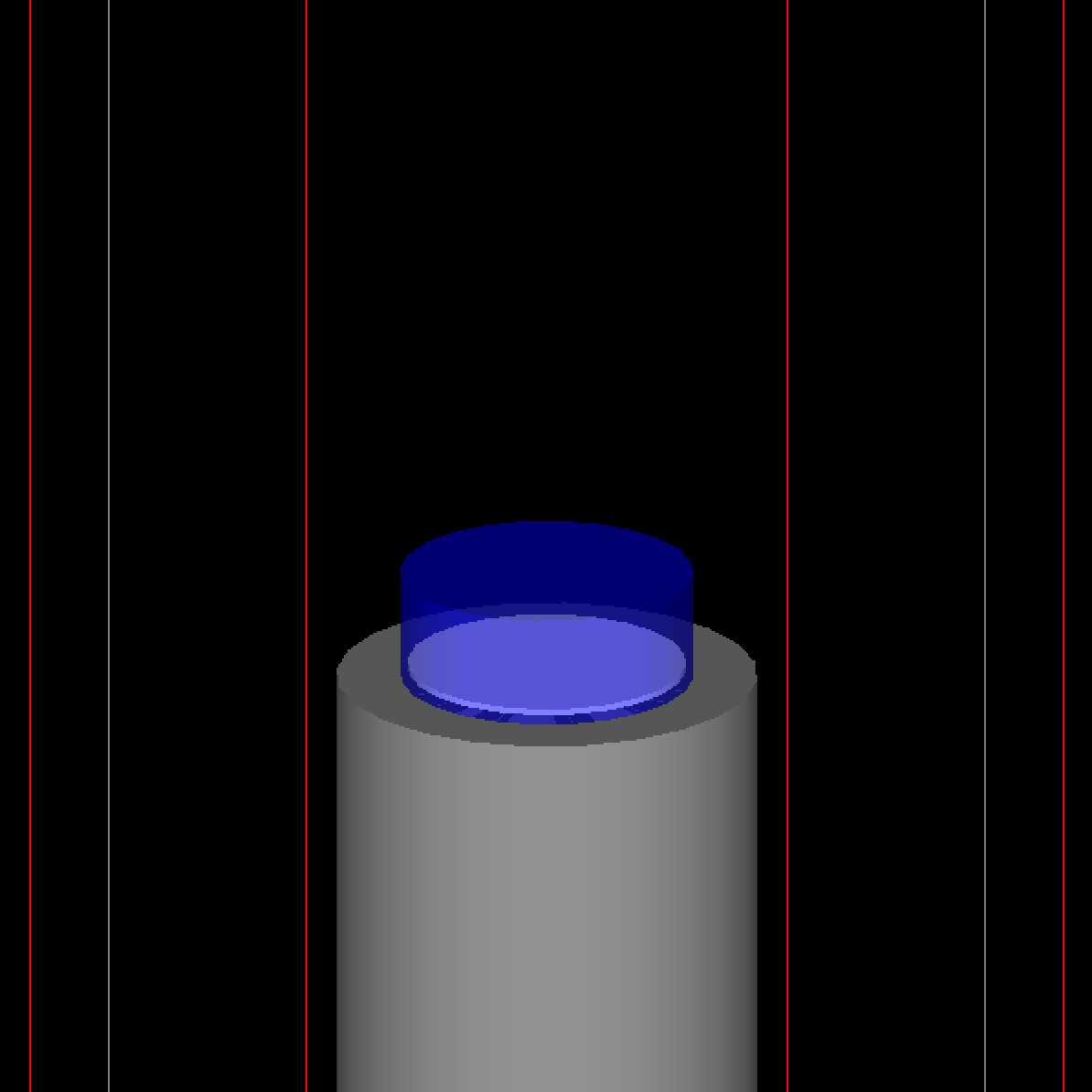} 
    \end{minipage}\hfill
    \begin{minipage}{0.45\textwidth}
        \centering
        \includegraphics[width=0.9\textwidth]{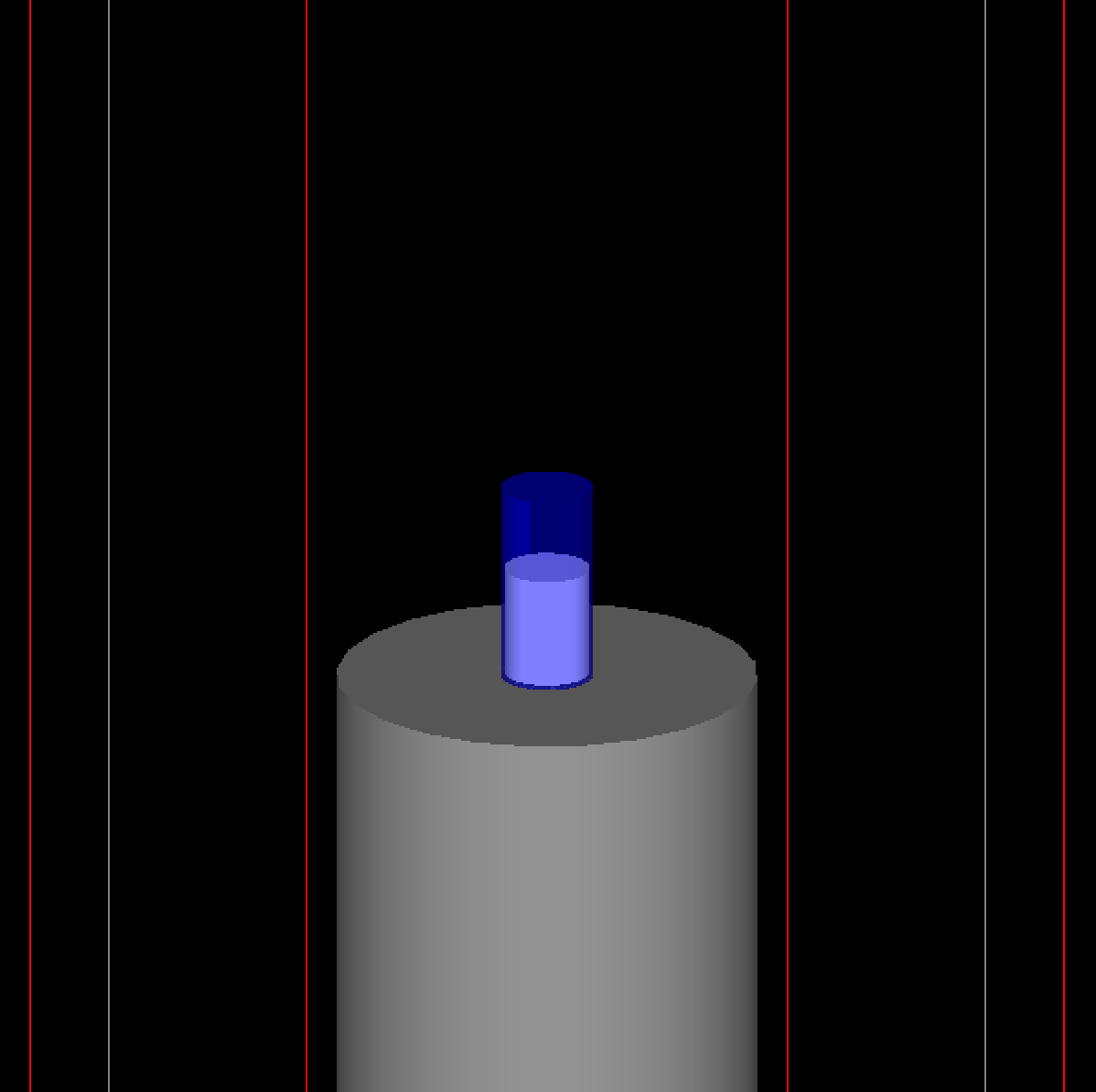} 
    \end{minipage}
    \caption{A \qty{9}{\g} powder sample in a petri dish (left) and a pot which is suitable for insertion in the well of Lumpsey (right). In this figure, both samples are shown sitting on the front face of the Roseberry detector.\label{fig:RoseberryPots}}
\end{figure}

For Roseberry and Lumpsey, a smaller sample of \qty{9}{\g} is used. For both Roseberry and Lumpsey, a pot suitable for insertion in the well is also simulated. However, this is a sub-optimal geometry for the planar BEGe detector. To compare optimised assay configuration for an identical sample, the Roseberry simulation is repeated but with the \qty{9}{\g} sample in a petri dish. Figure~\ref{fig:RoseberryPots} shows the two different configurations on the Roseberry detector.

\begin{figure}[ht!]
\includegraphics[width=\linewidth]{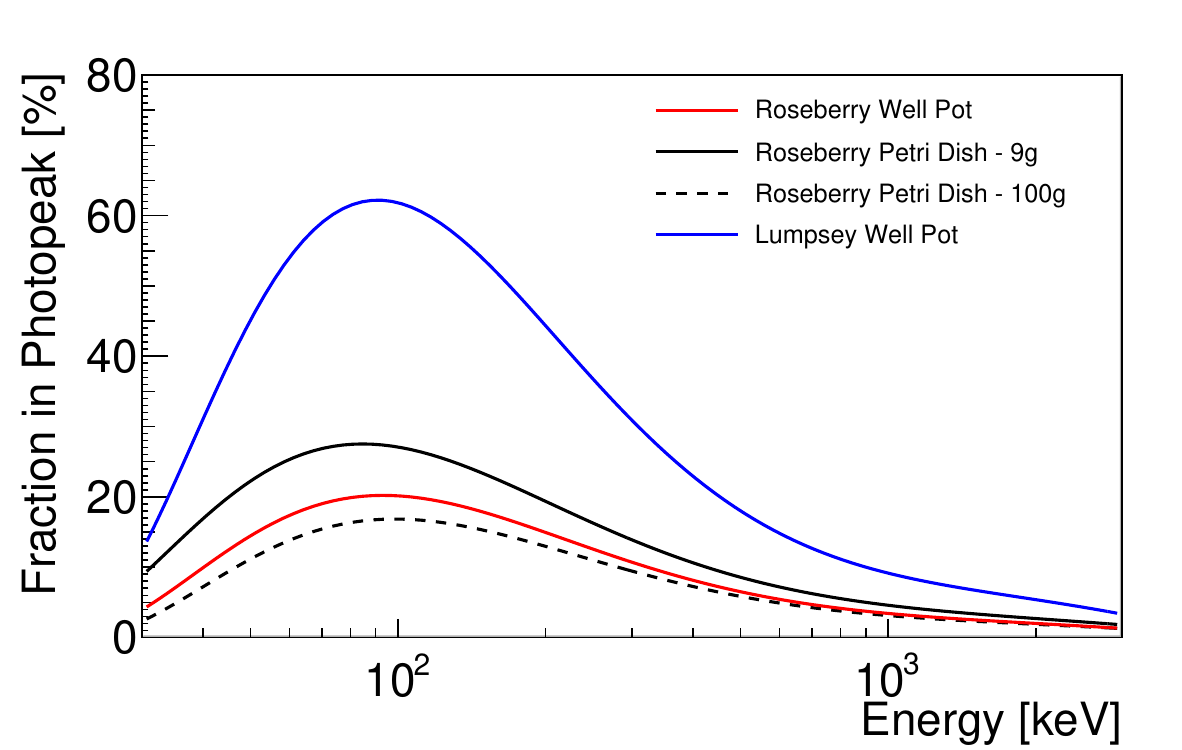}
\centering
\caption{\small{Comparison between the geometrical efficiency for a \qty{9}{\g} sample on Roseberry in a petri dish (black), and in a pot suitable for the well on Lumpsey (blue) and Roseberry (red). For comparison, the efficiency curve for the \qty{100}{\g} petri dish on Roseberry is shown (black dash). This shows that for a greater than 10 times increase in mass, the corresponding loss of efficiency is small.}}
\label{fig:efficiency}
\end{figure}

In Figure~\ref{fig:efficiency}, it can be seen that the highest efficiency overall for the \qty{9}{\g} sample is on Lumpsey. This figure also highlights that the petri dish sample on Roseberry gives a much better efficiency than does an identical mass sample in a well suitable pot.
With both geometric efficiency and background count rates determined, it is possible to predict the minimum detectable activity (MDA) for U/Th/K in each of the detectors. To calculate MDA for each radionuclide, we calculate the detection limit of number of counts required above background to give a \qty{90}{\percent} confidence level measurement. This is calculated using,

\begin{equation}
    L_D = k_{\alpha}^{2} + 2k_{\alpha} \sqrt{2B},
\end{equation}
where $L_{D}$ is the detection limit, $k_{\alpha}$ is the $k$-factor related to the error function of the Gaussian distribution ($k_{\alpha} = 1.282$ for a \qty{90}{\percent} confidence level) and $B$ is the number of counts expected from background sources over the duration of a planned radioassay.

From this, we calculate the corresponding sample-specific activity~\cite{Gilmore:2008abc,DONE201628} using,

\begin{equation}
    \textup{MDA} = \frac{L_{D}}{\varepsilon I_{\gamma}mt},
\end{equation}
where $\varepsilon$ is the energy dependent geometric efficiency of a specific sample, $I_{\gamma}$ is the gamma-ray intensity for the FEP being studied, $m$ is the mass of the sample and $t$ is the live time of the planned radioassay.

\begin{figure}[ht!]
\includegraphics[width=\linewidth]{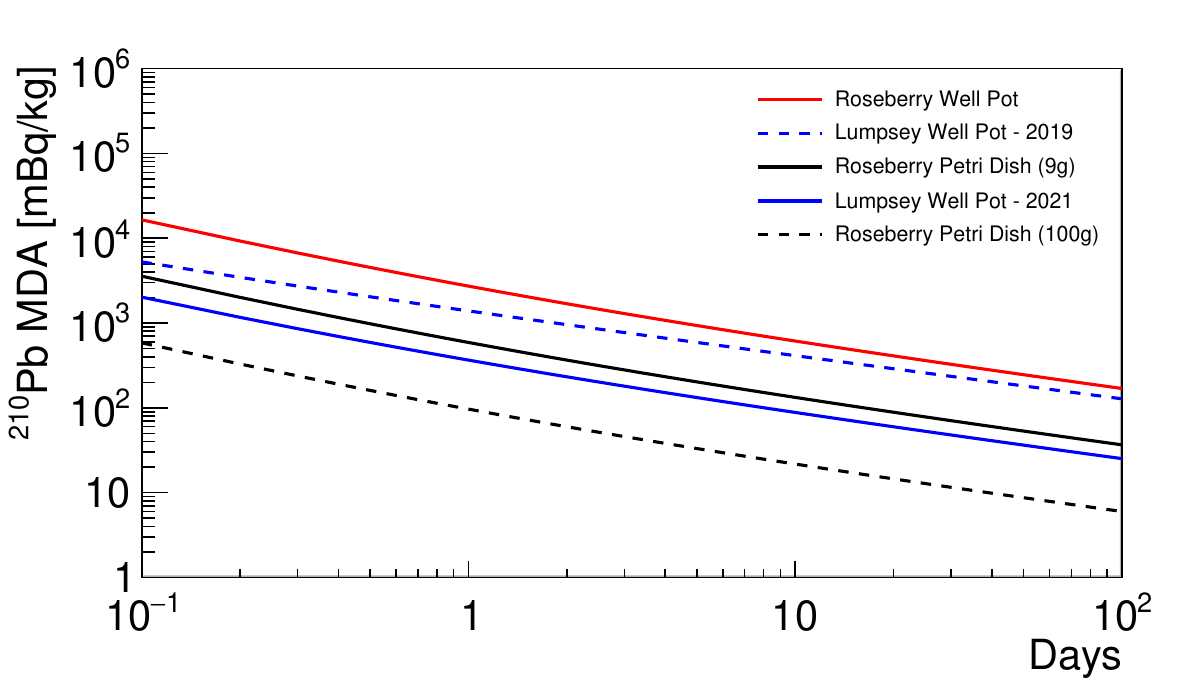}
\centering
\caption{\small{Comparison between the MDAs for $^{210}$Pb for a \qty{9}{\g} sample on Roseberry in a petri dish (black), and in a pot suitable for the well on Lumpsey (blue) and Roseberry (red). Lumpsey gives moderately lower MDAs for this \qty{9}{\g} sample however, Roseberry can improve on these MDAs for larger samples. The black dashed line is the MDA for a large petri dish sample on Roseberry. For comparison, the equivalent MDA for Lumpsey pre-refurbishment is shown by the blue dashed line. There is roughly a 5 times reduction in the MDA.}}
\label{fig:mda}
\end{figure}

\begin{table}[ht!]
\centering
\caption{Minimum detectable activities for each of the detectors in the BUGS facility. For Belmont, Merrybent and Lunehead the MDAs are for a \qty{4}{\litre} Marinelli beaker filled with \qty{4.5}{\kg} powder. For Roseberry and Chaloner, a petri dish containing \qty{100}{\g} of powder is used. For Lumpsey, a pot containing \qty{9}{\g} of powder is used. For $^{238}$U, $^{232}$Th and $^{40}$K, Lumpsey provides similar performance to Chaloner with a \qty{90}{\percent} smaller sample. \label{tab:mda}}

\begin{adjustbox}{width=\textwidth,center}
\begin{tabular}{|l|c|c|c|c|c|}
\hline
\multicolumn{1}{|c|}{\multirow{3}{*}{{Detector}}} &\multicolumn{1}{c|}{\multirow{3}{*}{{Mass (g)}}}& \multicolumn{4}{c|}{{MDA at 21 days}}\\
\cline{3-6}
&& {609 keV} & {238 keV} & {1461 keV} & {46.5 keV} \\
&& {$^{238}$U (\unit{\milli\becquerel\per\kilogram})} & {$^{232}$Th (\unit{\milli\becquerel\per\kilogram})} & {$^{40}$K (\unit{\milli\becquerel\per\kilogram})} & {$^{210}$Pb (\unit{\milli\becquerel\per\kilogram})} \\
\hline \hline
Belmont & 4500 & 0.2 & 0.09 & 1.8 & - \\
Merrybent & 4500 & 0.7 & 0.1 & 1.8 & - \\
Lunehead & 4500 & 1.0 & 0.9 & 8.4 & - \\
Roseberry & 100 & 2.5 & 1.9 & 36 & 18 \\
Chaloner & 100 & 12 & 5.1 & 119 & 28 \\
Lumpsey & 9 & 19 & 5.5 & 115 & 58 \\
\hline
\end{tabular}
\end{adjustbox}
\end{table}

\begin{table}[ht!]
\centering
\caption{Minimum detectable activities for the FEPs of interest in 0$\nu\beta\beta$ decay using the Belmont detector for a \qty{4.5}{\kg} Marinelli beaker sample. Assuming a long assay of \qty{100}{\day}, MDAs are at \qty{90}{\percent} confidence level and are presented in \unit{\micro\becquerel\per\kilogram} and \unit{\pico\g\per\g} (parts per trillion) of parent radionuclide equivalent (assuming secular equilibrium). Conversion from \unit{\micro\becquerel\per\kilogram} to \unit{\pico\g\per\g} is performed using the conversion factors detailed in~\cite{international1989iaea}.\label{tab:mda_0nubb}}
\begin{tabular}{|c|c|c|c|c|}
\hline
{Parent} & {Daughter} & {Energy} & \multicolumn{2}{c|}{{MDA at 100 days}}\\
\cline{4-5}
{Radionuclide} & {Radionuclide} & {(keV)} & {\unit{\micro\becquerel\per\kilogram}} & {\unit{\pico\g\per\g}  (U/Th)}\\
\hline
$^{238}$U & $^{214}$Pb & 351.6 & 70 & 5.7 \\
$^{238}$U & $^{214}$Bi & 609.3 & 101 & 8.2 \\
$^{238}$U & $^{214}$Bi & 1764.5 & 144 & 12 \\
$^{232}$Th & $^{212}$Pb & 238.6 & 38 & 9.4 \\
$^{232}$Th & $^{208}$Tl & 583.2 & 27 & 6.7 \\
$^{232}$Th & $^{208}$Tl & 2614.5 & 78 & 19 \\
\hline
\end{tabular}
\end{table}

Figure~\ref{fig:mda} shows the evolution of the MDAs with acquisition time for the $^{210}$Pb FEP at \qty{46.5}{\keV} for the \qty{9}{\g} sample in the three configurations on Roseberry and Lumpsey. It can be seen that both detectors are able to assay to similar sensitivities as long as the geometry of the sample used is optimised. It is also seen that there has been approximately a 5 times reduction in the MDA for $^{210}$Pb in Lumpsey when compared to pre-refurbishment. 

Table~\ref{tab:mda} shows the MDAs that could be achieved for the new detectors and highlights the gamma-ray peak for which this activity has been calculated. A combination of increased efficiency and substantially reduced background levels for the detectors presented here means these are far below what has been achieved using the original BUGS germanium detectors~\cite{Scovell:2017srl}. For Belmont and Merrybent, the MDAs for $^{238}$U, $^{232}$Th and $^{40}$K are shown and for Roseberry and Lumpsey, additionally the MDA for $^{210}$Pb is shown. For the \qty{609}{\keV} FEP, coincidence summing effects are calculated and applied as described in~\cite{Scovell:2017srl}. Belmont and Merrybent are both able to reach the equivalent of the \unit{\pico\g\per\g} (parts per trillion) level in U and Th and the \unit{\nano\g\per\g} (parts per billion) level in K. On first glance it may appear that Roseberry is less sensitive than Belmont but it is important to consider that the mass of the simulated sample on Roseberry is only \qty{2.2}{\percent} that of the sample on Belmont and the sample in the well of Lumpsey is only \qty{0.2}{\percent}. Thus Lumpsey and Roseberry could be affected by possible inhomogeneities in the distribution of the radioactivity inside the sample. 

Table~\ref{tab:mda_0nubb} shows the MDAs for radionuclides which can cause unwanted background for $0\nu\beta\beta$ radionuclides of interest. The measurement at \qty{351.2}{\keV} is for $^{214}$Pb but this can be used to infer the level of $^{214}$Bi assuming secular equilibrium. The calculated MDAs are comparable to those discussed in~\cite{Heusser:2006hgf} showing that the BUGS HPGe detectors are approaching the current world leading sensitivities. 

\section{Conclusions}
Germanium detectors of ever increasing sensitivity to samples of all shapes and sizes are going to be of key importance to material radioassay for next generation ultra-low background particle physics experiments. The S-ULB detectors operating in the BUGS facility aim to meet the needs of these experiments by providing high sensitivity measurements that can span the entire energy spectrum of interest to include measurements of the $^{210}$Pb FEP at \qty{46.5}{\keV} all the way up to $^{208}$Tl at \qty{2615}{\keV}. For typical samples, minimum detectable activities of \qty{<1}{\milli\becquerel\per\kg}, \qty{<0.1}{\milli\becquerel\per\kg} and \qty{<10}{\milli\becquerel\per\kg} can be reached after 21 days on the Belmont detector for $^{238}$U, $^{232}$Th and $^{40}$K, respectively. The Roseberry detector is able to reach \qty{<18}{\milli\becquerel\per\kg} for $^{210}$Pb over the same period. This increase in sensitivity, in turn, will improve throughput for radioassay, allowing materials to be characterised faster than on the original BUGS detectors.

To improve confidence in the measurements taken using the BUGS detectors, regular calibrations are performed to monitor the stability of the detector response. Additionally, data is acquired such that any periods of instability can be easily identified and removed, such as when the ambient radon levels in the underground laboratory fluctuate.

\section{Acknowledgements}
This work has been supported by the Science and Technology Facilities Council.
The authors would like to acknowledge our hosts ICL-Boulby who have been of great support throughout the development of the Boulby Underground Laboratory and of BUGS.
The University of Edinburgh is a charitable body, registered in Scotland, with the registration number SC005336.


 \bibliographystyle{JHEP}
 \bibliography{main.bib}

\end{document}